**Article type: Full Paper**

**Titanium oxynitride thin films with tunable double epsilon-near-zero behaviour**

*Laurentiu Braic, Nikolaos Vasilantonakis, Andrei P. Mihai\*, Ignacio J. Villar Garcia, Sarah Fearn, Bin Zou, Brock Doiron, Rupert F. Oulton, Lesley Cohen, Stefan A. Maier, Neil McN. Alford, Anatoly V. Zayats & Peter K. Petrov*

Dr. Laurentiu Braic
National Institute for Optoelectronics, 409 Atomistilor, 077125, Magurele, Romania
Dr A. P. Mihai, Dr. Bin Zou, Dr. Sarah Fearn, Dr. Ignacio J. Villar Garcia, Prof. N. Alford, Dr P. K. Petrov
Imperial College London, Department of Materials, Prince Consort Road, London SW7 2BP, UK
E-mail: a.mihai@imperial.ac.uk; p.petrov@imperial.ac.uk
B. Doiron, Dr R. F. M. Oulton, Prof Lesley Cohen, Prof. S. A. Maier
Imperial College London, Department of Physics, Prince Consort Road, London SW7 2AZ, UK
Dr. Nikolaos Vasilantonakis, Prof. Anatoly V. Zayats
King's College London, Department of Physics, Strand, London WC2R 2LS, UK

Titanium Oxynitride (TiOxNy) thin films are fabricated using reactive magnetron sputtering. The mechanism of their growth formation is explained and their optical properties are presented. The films grown when the level of residual Oxygen in the background vacuum was between 5E-9Torr to 20E-9Torr exhibit double Epsilon-Near-Zero (2-ENZ) behaviour with ENZ1 and ENZ2 wavelengths tunable in the 700-850 nm and in the 1100-1350 nm spectral ranges, respectively. Samples fabricated when the level of residual Oxygen in the background vacuum was above 2E-8Torr exhibit non-metallic behaviour, while the layers deposited when the level of residual Oxygen in the background vacuum was below 5E-9Torr, show metallic behaviour with a single ENZ value. The double ENZ phenomenon is related to the level of residual

Oxygen in the background vacuum and is attributed to the mixture of TiN and TiOxNy/TiOx phases in the films. Varying the partial pressure of nitrogen during the deposition can further control the amount of TiN, TiOx and TiOxNy compounds in the films and, therefore, tune the screened plasma wavelength. A good approximation of the ellipsometric behaviour is achieved with Maxwell-Garnett theory for a composite film formed by a mixture of TiO2 and TiN phases suggesting that double ENZ TiOxNy films are formed by inclusions of TiN within a TiO2 matrix. These oxynitride compounds could be considered as new materials exhibiting double ENZ in the visible and near-IR spectral ranges.

## 1. Introduction

As optoelectronic components become nano-dimensional, controlling the coupling between light and matter at the nanoscale has become a major technological challenge, as well as the subject of theoretical studies. Plasmonics has been established as the backbone of this new nanophotonic technology, exploiting the strong interactions between electromagnetic radiation and plasma oscillations in metallic nanostructures, which lead to the establishment of hybrid modes overcoming the diffraction limit[1,2]. Traditionally based on materials such as Ag and Au, it is now recognized that the materials base should be widened with a view to integration with semiconductor (e.g., CMOS-based) technology.

In the search for alternative plasmonic materials, titanium nitride (TiN) was proposed as a possible candidate to be used in the near infrared domain[3,4]. TiN films are one of the most studied systems, finding widespread use in a number of applications[5,6,7,8,9]. The suitability of TiN films for plasmonic application has been recently the subject of extended research[10,11]. The colour of stoichiometric TiN films[12,13] hints at their similarity to gold. As confirmed by ellipsometric measurements[14, 15], the plasma frequency in recent plasmonics literature is ~500 nm and the epsilon-near-zero (ENZ) behaviour (where the real part of permittivity is close to zero), extends up to around 1000 nm. In turn, its counterpart Titanium dioxide (TiO2) is a well-known insulator and oxide semiconductor with many applications in important research areas such as solar energy harvesting[16] and photocatalysis[17].

A number of intermediate phases of general composition TiOxNy called Titanium oxynitrides are found midway between TiN and TiO2. Their properties are similar to those of the respective parent materials when the composition is near that of pure systems. They change in properties when they evolve from one pure phase to the other. However, there are several questions about these intermediate oxynitride phases that still need to be answered. In particular, whether they are stable and how their optical properties depend on the TiOxNy composition. Preparation of Titanium oxynitrides through either oxidation of TiN or nitridation of TiO2 is not straightforward. On one hand, the oxidation of the TiN quickly leads to formation of TiO2[18] and on the other, nitrogen implantation in TiO2 leads to substantial reconstruction of the surface due to strong reduction and only a small amount (2-3%) of implanted nitrogen can be reached[19].

A major breakthrough, in terms of plasmonic applications, has been the development of ENZ metamaterials in the desired frequency range, which have many anticipated applications[20] as a result of their perfect absorber behaviour, supercoupling effect[21,22], radiation directivity[23,24], and nonlinearity enhancement[25, 26]. Recently, a broadband ENZ in the near-infrared has been reported for a multilayer of Indium Tin Oxide (ITO) films having different doping densities[27] as well as for nanorod-based metamaterials[28]. Another option to achieve broadband ENZ has been the addition of metallic inclusions to a dielectric film, with the filling factor changing along the direction of growth[29]. Metal-dielectric layers of varying thicknesses and geometries, and transparent conductive oxides, have also been suggested[24,30].

Here, the mechanism of formation of titanium oxynitride thin films and their optical properties with tunable ENZ behaviour is investigated. We present the technological conditions for deposition of titanium oxynitride thin films with unusual double ENZ frequencies. It is shown that they can be modified by changing the film deposition conditions. We propose these oxynitrides as alternative plasmonic materials exhibiting double ENZ in the visible and near-IR spectral ranges.

## 2. Results

A series of TiOxNy thin films have been deposited on MgO and Si substrates by reactive sputtering of Ti targets at room temperature using an Ar-N2 gas mixture. The sputtering system was equipped with Residual Gas Analyser (SRS), which was used to measure amount of residual gases in the vacuum chamber before deposition, and therefore to control the level of background Oxygen by pre-deposition chamber conditioning (e.g., high temperature baking of the and/or pre-sputtering of titanium, which is known as a good oxygen getter).

In the first set of experiments, a series of 50 nm TiOxNy films were sputtered on Si substrates using a gas mixture of 70%Ar+30%N2 (the total deposition gas pressure was 1.5 mTorr). A Residual Gas Analysis of the vacuum was carried out before starting the growth. Three different scenarios were studied: first deposition at room temperature, without any chamber preparation (baking or pre-sputtering of Titanium); second, prior deposition at room temperature, the chamber and substrates were baked at 250°C for 2 hours; and third the chamber and the substrate were heated to 600°C and kept at this temperature for 30 minutes while we performed a Ti pre-sputtering. The effect of the chamber preparation on the residual oxygen in the chamber can be seen in the **Table 1**.

The optical properties presented in **Figure 1** (a and b), show the different behaviour of the real and imaginary parts of the dielectric constant of the deposited thin films obtained from ellipsometric measurements. When comparing these behaviours with the Residual Gas Analysis data (presented in Table 1), one can draw the conclusion that the reason for the different optical properties is linked to the level of residual Oxygen in the background vacuum.

Samples fabricated when the level of residual Oxygen in the background vacuum was above 20 nTorr (High Residual Oxygen) at both room and high temperatures exhibited non-metallic (TiO2 – like) behaviour, while the layers deposited when the level of residual Oxygen in the background vacuum was below 5 nTorr (Low Residual Oxygen), showed metallic (TiN – like) behaviour with ENZ value around a wavelength of 600 (Figure 1a). The real part of their dielectric constants Re{ε} is continuously decreasing, with negative values throughout most of the visible and infrared ranges, while the imaginary part constantly increases almost linearly. This behaviour corresponds to what has already been reported for TiN films grown at high temperatures[10].

The films deposited with the level of residual Oxygen in the background vacuum between 5 nTorr to 20 nTorr (Medium Residual Oxygen) yield a 2ENZ behaviour with 2 ENZ frequencies around 600 nm and 1500 nm. This unusual phenomenon is not influenced by the films' crystallinity. We attribute the appearance of double ENZ frequency to the mixture of TiN and TiOxNy/TiOx phases in the films.

To better understand the influence of the mixing of the TiN and TiOx phases on the optical properties of the Medium Residual Oxygen deposited samples, a Maxwell- Garnett formalism has been used to simulate the experimentally observed optical properties (Figure 1c and d). In this model, the oxinitrite films were treated as a composite of different phases and effective dielectric constants were calculated. Two scenarios were considered with either TiO2 inclusions in TiN films or, the other way around, TiN inclusions in TiO2 films. We used literature data for TiN and TiO2[38] and the Maxwell Garnett model[39]. The simulation results suggest that the double ENZ behaviour ellipsometry appears only for the case of TiN inclusions into the TiO2 matrix, while TiO2 inclusions in the TiN matrix do not result in double-ENZ behaviour. The latter can be easily understood as TiN inclusions would have a plasmonic polarizability while TiO2 inclusions would not. It should be noted also that in both cases, the presence of TiO2 for high TiN content appears to red shift the ENZ point. Therefore, even the Low Residual Oxygen deposited films with $\varepsilon=0$ near 700 nm could well include TiO2 (which is confirmed by the XPS and SIMS results and will be discussed below). For pure TiN films, the ENZ is expected to be close to 500 nm. These results of the Maxwell - Garnett modelling correspond remarkably well to our experimental results confirming the formation mechanism of TiOxNy films as TiN inclusions into a TiO2 matrix.

The partial pressure of N2 (within the N2-Ar sputter gas mixture) (as well as the deposition temperature) govern the nitride saturation of the growing TiN film. Reducing the amount of nitrogen in the sputter gas mixture, results in the formation of non-saturated TiN layer, which forms TiOx and TiOxNy phases due to residual oxygen inside the growth chamber.

Varying the content of the deposition gas mixture, one can control the amount of TiN, TiOx and TiOxNy in these films and therefore change the optical properties of the samples (see Figure 2, discussed in detail below).

To explore the above hypothesis a second set of experiments was carried out. A series of thin films were sputtered at room temperature in a chamber with Medium Residual Oxygen level using sputter gas mixture with N2 partial pressure varying in the range between 5% and 30% of the total gas pressure. The samples' optical measurement results are presented in **Figure 2**.

These films exhibit two distinct epsilon-near-zero frequencies (Figure 2a). The real part of the dielectric constant of the samples deposited at room temperature and nitrogen partial pressures between 5% and 30% decreases with the increase of wavelength from the ultra-violet range, becoming negative around 700—750 nm, and starting to increase around 1000-1100 nm and finally becoming positive again after 1350 nm. For the samples fabricated at nitrogen partial pressures of 5%, the overall behaviour is the same, but the two ENZ frequencies are 850 nm and 1200 nm. The imaginary part of these samples' dielectric constant (Figure 1b) is nearly constant until 600 nm, and then significantly increases throughout the measured range showing losses, significantly higher than values reported for Au films[10] at the 2$^{nd}$ ENZ point. We attribute the tunable double ENZ behaviour of the room temperature deposited samples to the mixture of TiN and TiOxNy/TiOx phases in the films. Moreover, the XPS studies show that for the Medium Residual Oxygen fabricated films, the ratio between TiN and TiOx does not depend on the depth, but we observed higher oxidation near the surface (which is normal after exposure to air).

Disappearance of the double-ENZ behaviour in the samples deposited at high temperature is due to the (nearly) complete TiN structure saturation caused by the temperature enhanced nitrogen diffusion and low reactivity with available oxygen.

To verify this hypothesis, we compared the Low Residual Oxygen deposited samples (grown at 600°C) with the Medium Residual Oxygen deposited samples (grown at room temperature). Samples were placed in the XPS system and in-situ Ar ion milled for 20 seconds before an XPS spectra were acquired. The process of milling-data acquiring was repeated in cycles for up to 300 seconds total sputtering time. The ion sputtering rate has been estimated to be approx.0.26nm/min. The XPS results become constant (i.e., surveys stopped changing) after the third milling- data acquiring cycle. **Figure 3** shows the Ti 2p (Figure 3a) and N 1s (Figure 3b) XPS spectra of samples deposited at 20% N2 partial pressure in vacuum chamber with Low Residual Oxygen, and Medium Residual Oxygen. The spectra show signals at the characteristic binding energies of TiNx, TiOxNy, and TiOx[31, 32]. Table 2 shows the calculated relative atomic percentages of the three different chemical states of Ti in the samples (the fitting procedures used to calculate these percentages are shown in the supplementary information). Low Residual

Oxygen deposited samples consist mainly of titanium nitride, with a small amount of titanium dioxide (3.5%) and oxynitride (9.6%). Medium Residual Oxygen deposited samples exhibit a considerably higher content of both titanium dioxide (10.1%) and titanium oxynitride (18.6%). These results are confirmed by Secondary Ion Mass Spectroscopy (SIMS) measurements on 20% N2 partial pressure samples (Figure 3c and d). SIMS depth profiling shows that the film composition is uniform, except for the enhanced secondary ion signals near the surface (no capping layer has been used) and at the interface with the MgO substrates.

The higher content of oxidation states of Ti in the Medium Residual Oxygen fabricated samples, is due to the greater proportion of broken-bond defects or unreacted Ti atoms, to which residual oxygen then becomes attached. The N 1s peak at ~397 eV for all the films (see supplementary figure 13S) in the present case can be attributed to nitrogen in the Ti-N bonds[33]. The very intense nature of this N 1s component peak is indicative of nitride formation in the film. The N 1s peak at lower binding energy (~396 eV) has been associated to terminally bound nitrogen that is released during nitridation of Ti-O sites or in surface oxynitrides[32,33]. This peak has been noted by several authors for nitrides and oxynitrides of Ti grown by various methods[33, 34]. In fact, in a theoretical paper, Graciani et al.[35] argue that the most stable way of growing good quality Ti oxynitrides is through nitrogen incorporation of the most stable alpha-TiO phase, which is isostructural with TiN. According to their calculations, it is much easier to implant N2 into the TiO than implant N2 into TiO2, which is the usual way of growing TiOxNy[33]. Our electrical measurements show the resistivity of the samples decreases as the intensity of the XPS TiN characteristic signal increases in comparison to the intensity of the characteristic TiOxNy signal ; the characteristic TiN signal at approximately 455 eV increases in comparison with the TiO2 and TiOxNy signals in the Ti 2p spectra (see Figure 3 and Table S1 in the supplementary information). A full survey of the relative percentages of each of the Ti, O and N components found in the films can be found in the supplementary data.

The films' electronic properties were studied using Kelvin Force Microscopy (KFM) and Hall effect measurements. To check the electrical film uniformity, we performed KFM on both Low and Medium Residual Oxygen grown films (results in supplementary) and found a complete uniformity in the electrical conductivity, with no microscopic segregated phases.

All samples grown at Medium Residual Oxygen exhibited electron concentrations in the order of $10^{22}$ carriers/cm3 (**Figure 4a**), similar to the values encountered in the literature for titanium nitride and oxynitride[31]. The samples' electron mobility (Figure 4b) increased from 0.03 cm2V-1s-1, for samples fabricated at 5% nitrogen partial pressure, to 0.1 cm2V-1s-1 for samples fabricated at 30% nitrogen partial pressure, respectively. This is in agreement with earlier results for samples with similar Ti:O ratios[31]. Furthermore, such small mobility values are to be expected for sputtered polycrystalline films.

Very recently, two studies regarding the effect of post growth vacuum annealing on RF sputtered TiN films[36], and the effect of changing the deposition conditions (gas mixture, substrate temperature, biasing of substrate during growth)[37] have shown a wide tunability of the screened plasma wavelength in the 500—720 nm spectral range. This was attributed to a change in the crystallinity or composition with annealing temperature, but no double ENZ behaviour has been reported.

The optical losses observed in the samples grown at Medium and High Residual Oxygen (Figure 1b and Figure 2b) are like the optical losses of the TiN films, and realtively higher than the losses usually measured in Transparent Conductive Oxides (TCO) such as ITO, AZO and GZO[3]. This is not surprising given that contributions to losses come from both TiN and $TiO_2$/$TiO_xN_y$ phases found in these films. TCOs are known to provide low carrier concentrations which in turn lead to a low imaginary dielectric permittivity, and hence low losses. Transition metal nitrides have higher losses because of their large carrier concentration. That is the case in the comparison in Figure 1b where the low residual oxygen sample has the highest losses compared to the RT high residual oxygen sample. However, at the 2nd ENZ point the losses reported in our medium residual oxygen grown samples are even higher than those found at the first ENZ, leading us to believe that even though the sample become dielectric in its optical behaviour, the TiN contribution (compared to the $TiO_2/TiO_xN_y$) is still important enough to produce a large net carrier concentration (see Figure 4 a) and hence high imaginary dielectric permittivity. Finally, despite its relatively higher (compared to traditional TCOs) optical losses in the near IR spectre, TiON is attractive for plasmonic applications mainly because its controlled ENZs tunability, low cost, and high temperature stability and easy integration with the semiconductor technology.

## 4. Conclusion

We have shown that composite films containing nanocrystalline phases of Titanium nitride and Titanium oxide exhibit unusual double ENZ properties that can be tuned via controlling deposition parameters and thus influencing the ratio between these phases. The reason for its appearance is the level of residual Oxygen in the background vacuum that reacts with Ti during deposition. We have also confirmed that the double-ENZ behaviour appears only on the samples where TiN is incorporated in pre-existing TiO2 matrix. The two ENZ frequencies of the TiOxNy films could be further tuned by changing the partial pressure of the nitrogen in the sputter gas mixture. We should emphasize that in our studies the double ENZ behaviour has been observed, in both thin amorphous TiOxNy films on MgO and Si substrates, thus allowing one to fabricate, control and engineer tunable plasmonic and metamaterial devices, using CMOS compatible technology.

## 5. Experimental

Titanium (oxy)nitride (TiOxNy) thin films were deposited on 10 x 10 mm2 MgO (001) substrates (Crystal GmbH Berlin) and on Si(100) p-doped substrates, by RF reactive magnetron sputtering, from a high purity Ti (99.995%) target (Pi-kem), using a MANTIS Deposition

System. The power fed on the Ti cathode was maintained constant (200 W). The deposition was carried out using a gas mixture of Ar and N2 with total constant pressure of (0.2 Pa). The samples have been fabricated at nitrogen partial pressures (N2/(Ar+N2)) of 5, 10, 20 and 30% at two different temperatures: room temperature and 600 0C. Nominal thicknesses of 50 nm were targeted but a thickness of around 40±5 nm was obtained for higher N2 content. (This is a known effect because of target nitridation [32] and corresponding lower deposition rate). All films were analyzed ex-situ after at least 96 hours exposure to clean air, at room temperature.

X-Ray Diffraction (XRD) has shown that the films deposited at RT were amorphous, while the film deposited at 600 0C exhibited a textured structure with (001) orientation (not shown here). A Phillips Panalytical MPD X-ray diffraction system, equipped with a graphite monochromator coupled with a scintillation counter detector, using Cu Kα radiation (λ = 1.5405 nm) working in the Bragg-Brentano was used to investigate the crystallographic structure and phase composition of the films. Atomic force microscopy (AFM) was performed using a Bruker INNOVA instrument. The film thickness of all samples measured using a Dektak 150 surface profiler.

All XP spectra were recorded using a K-alpha+ XPS spectrometer equipped with a MXR3 Al Kα monochromated X-ray source (hν = 1486.6 eV). X-ray gun power was set to 72 W (6 mA and 12 kV). With this X-ray setting, the intensity of the Ag 3d5/2 photoemission peak for an atomically clean Ag sample, recorded at a pass energy (PE) of 20 eV, was $5 \times 10^6$ counts s-1 and the full width at half maximum (FWHM) was 0.58 eV. Binding energy calibration was made using Au 4f7/2 (83.96 eV), Ag 3d5/2 (368.21 eV) and Cu 2p3/2 (932.62 eV). Charge compensation was achieved using the FG03 flood gun using a combination of low energy electrons and the ion flood source. Argon milling of the samples was done using the standard EX06 Argon ion source using 500 V accelerating voltage and 1 µA ion gun current. Survey scans were acquired using 200 eV pass energy, 0.5 eV step size and 100 milliseconds (50 ms x 2 scans) dwell times. All high-resolution spectra (C 1s, N 1s, Ti 2p and O 1s) were acquired using 20 eV pass energy, 0.1 eV step size and 1 second (50ms x 20 scans = 1000 ms) dwell times. For these measurements, samples were prepared by pressing the sample onto carbon based double side sticky based tape and analysed at an electron take-off angle normal to the surface with respect to the analyser.

Casa XPS was used for data interpretation. Shirley or two-point linear background subtractions were employed depending on background shape. Thermo scientific sensitivity factors were used for quantification analysis. Peaks were fitted using GL(30) line shapes; a combination of a Gaussian (70%) and Lorentzian (30%). All XP spectra were charge corrected by referencing to the $TiO_2$ peak in the Ti 2p signal to 458.5 eV.

SIMS depth profiling of the thin films was carried out using an IONTOF ToF-SIMS 5 instrument. The analytical ion beam used was a 25 keV $Bi^+$ LMIG, in high current bunch mode with a beam current of approximately 1 pA. The analytical area was 100 µm2. Depth profiling was performed with a 1 keV $Cs^+$ ion beam with a current of 75 nA and the sputter crater area was 300 ◻m2. Both positive and negative secondary ions were collected.

The electrical resistivity, charge carrier concentration and mobility of the films were investigated by using a Hall measurement system in Van der Pauw geometry at 1.6 T magnetic field, alternated at 50 mHz, using soldered indium contacts.

For the optical characterisation of the films a Horiba Jobin-Yvon Uvisel 2 ellipsometer was employed. Phase modulation reflection ellipsometry measurements were carried out at a 70° angle of incidence, with an optical spot size of 2 x 0.7 mm2. As it is the case when the films thickness is much smaller than the incident wavelength, a three-phase model was used, for the two interfaces: air/film and film/substrate. The optical constants of the films were extracted from the acquired spectra using the Marquardt minimization algorithm; the optical constants of the MgO substrate and the thickness of the TiOxNy films were considered to be known; therefore, the optical constants of the films were fitted to the experimental data, without considering any pre-defined function or theoretical model.

**Acknowledgements**

All authors acknowledge the EPSRC Reactive Plasmonics Programme Grant EP/M013812/1. S.A.M. further acknowledges the Royal Society and the Lee-Lucas Chair for funding. A.V.Z. acknowledges support from the Royal Society and Wolfson Foundation.
.

Received: ((will be filled in by the editorial staff))

Revised: ((will be filled in by the editorial staff))

Published online: ((will be filled in by the editorial staff))

1. Giannini, V., Fernández-Domínguez, A. I., Heck, S. C., & Maier, S. A. Plasmonic nanoantennas: fundamentals and their use in controlling the radiative properties of nanoemitters. *Chemical Reviews*, **111**(6), 3888-3912 (2011).

2. Barnes, W. L., Dereux, A., & Ebbesen, T. W. Surface plasmon subwavelength optics. *Nature*, **424**(6950), 824-830 (2003).

3. Naik, G., Kim, J. and Boltasseva, A. Oxides and nitrides as alternative plasmonic materials in the optical range, *Opt. Mater. Express* **1**, 1090–1099 (2011).

4. Naik, G. V., Shalaev, V. M. and Boltasseva, A. Alternative plasmonic materials: beyond gold and silver, *Advanced Materials*, vol. 25, **no. 24**, 3264-3294 (2013).

5. Perry, A.J. A contribution to the study of Poisson's ratios and elastic constants of TiN, ZrN and HfN. *Thin Solid Films*, **193–194**, 463–471 (1990).

6. Musil, J. Hard and superhard nanocomposite coatings, *Surface and Coatings Technology*, Volume **125**, Issues 1–3, , 322-330 (2000).

7. Patsalas, P., Charitidis, C. and Logothetidis, S. The effect of substrate temperature and biasing on the mechanical properties and structure of sputtered titanium nitride thin films, *Surface and Coatings Technology*, Volume **125**, Issues 1–3, 335-340 (2000).

8. R. Leutenecker, B. Fröschle, U. Cao-Minh, P. Ramm, Titanium nitride films for barrier applications produced by rapid thermal CVD and subsequent in-situ annealing, *Thin Solid Films*, Volume **270**, Issues 1–2, 621-626, (1995).

9. Liu, Y.X. et al., Fin-height controlled TiN-gate FinFET CMOS based on experimental mobility, *Microelectronic Engineering*, Volume **84**, Issues 9–10, Pages 2101-2104 (2007).

10. Naik, G., Schroeder, J.L. , Ni, X., Kildishev, A.V., Sands, T.D., Boltasseva, A. Titanium nitride as a plasmonic material for visible and near-infrared wavelengths, *Optical Material Express*, vol. **2**, no. 4, pp. 478-489 (2012).

11. Patsalas, P., Kalfagiannis, N., Kassavetis, S. Optical properties of Plasmonic performance of Titanium Nitride, *Materials*, **8**, 3128 − 3154 (2015).

12. Petrov, I., Barna, P.B., Hultman, L., Greene, J.E. Microstructural evolution during film growth. *J. Vac. Sci. Technol. A*, **21**, S117–S128 (2003).

13. Niyomsoan, S., Grant, W., Olson, D.L., Mishra, B. Variation of color in titanium and zirconium nitride decorative thin films, *Thin Solid Films*, **415**, Issues 1–2, 187-194 (2002).

14. Logothetidis, S., Alexandrou, I., Stoemenos, J. In-situ spectroscopic ellipsometry to control the growth of Ti nitride and carbide thin films, *Applied Surface Science*, Volume **86**, Issues 1–4, 185-189 (1995).

15. Patsalas, P. and. Logothetidis, S Optical, electronic, and transport properties of nanocrystalline titanium nitride thin films, *J. Appl. Phys*. **90**, 4725–4734 (2001).

16. Grätzel, M. Photoelectrochemical cells, *Nature* **414**, 338 (2001).

17. Asahi, R., Morikawa, T., Ohwaki, T., K. Aoki, and Taga, Y., *Science* **293**, 269 (2001).

18. J. Graciani, J. F. Sanz, T. Asaki, K. Nakamura, and, J. A. Rodríguez, *J. Chem. Phys.* **126**, 244713 (2007).

19. M. Batzill, E. Morales, and U. Diebold, *Phys. Rev. Lett*. **96**, 026103 (2006).

20. Silveirinha, M. G., Alù, A., Edwards, B., & Engheta, N. Overview of theory and applications of epsilon-near-zero materials. In *URSI General Assembly*, (2008).

21. Silveirinha, M., Engheta, N. Tunneling of Electromagnetic Energy through Subwavelength Channels and Bends using ε-Near-Zero Materials, *Phys. Rev. Lett*. **97**, 157403 (2006).

22. Edwards, B., Alù, A., Young, M., Silveirinha, M., Engheta, N. Experimental Verification of ε-Near-Zero Metamaterial Supercoupling and Energy Squeezing Using a Microwave Waveguide, *Phys. Rev. Lett.* **100**, 033903, (2008).

23. Ziolkowski, R. W. Propagation in and scattering from a matched metamaterial having a zero index of refraction, *Phys. Rev. E* **70**, 046608 (2004).

24. Braic, L., Vasilantonakis, N., Zou, B., Maier, S. A., Alford, N. M., Zayats, A. V., & Petrov, P. K.. Optimizing Strontium Ruthenate Thin Films for Near-Infrared Plasmonic Applications. *Scientific reports*, **5**, (2015).

25. Wurtz, G. A. *et al.* Designed ultrafast optical nonlinearity in a plasmonic nanorod metamaterial enhanced by nonlocality. *Nature nanotechnology*, **6**(2), 107-111, (2011).

26. Neira, Andres D., et al. Eliminating material constraints for nonlinearity with plasmonic metamaterials, *Nature communications* **6** (2015).

27. Yoon, Junho, et al. Broadband Epsilon-Near-Zero Perfect Absorption in the Near-Infrared., *Scientific reports* **5** (2015).

28. Nasir, M. E., Peruch, S., Vasilantonakis, N., Wardley, W. P., Dickson, W., Wurtz, G. A., & Zayats, A. V. Tuning the effective plasma frequency of nanorod metamaterials from visible to telecom wavelengths. *Applied Physics Letters*, **107**(12), 121110, (2015).

29. Sun, L., Yu, K.W. Strategy for designing broadband epsilon-near-zero materials, *J. Opt. Soc. Am. B* / Vol. **29**, No. 5 (2012).

30. Goncharenko, Anatoliy V., and Kuan-Ren Chen. Strategy for designing epsilon-near-zero nanostructured metamaterials over a frequency range, *Journal of Nanophotonics* **4.1**, 041530-041530, (2010).

31. Von Seefeld, Hermann, et al. Investigation of titanium—nitride layers for solar-cell contacts. Electron Devices, *IEEE Transactions* on **27.4**, 873-876 (1980).

32. Saha, Naresh C. and Tompkins, Harland G. Titanium nitride oxidation chemistry: An x-ray photoelectron spectroscopy study, *Journal of Applied Physics*, **72**, 3072-3079 (1992).

33. Drygas, M., Czosnek, C., Paine, R. T., & Janik, J. F. Two-stage aerosol synthesis of titanium nitride TiN and titanium oxynitride TiO x N y nanopowders of spherical particle morphology. *Chemistry of materials*, **18**(13), 3122-3129 (2006).

34. Jouan, P. Y., Peignon, M. C., Cardinaud, C., & Lemperiere, G. Characterisation of TiN coatings and of the TiN/Si interface by X-ray photoelectron spectroscopy and Auger electron spectroscopy. *Applied Surface Science*, **68**(4), 595-603 (1993).

35. Graciani, J., Hamad, S., & Sanz, J. F. Changing the physical and chemical properties of titanium oxynitrides TiN 1− x O x by changing the composition. *Physical Review B*, **80**(18), 184112 (2009).

36. Wang, Y., Capretti, A., & Dal Negro, L. Wide tuning of the optical and structural properties of alternative plasmonic materials. Optical Materials Express, 5(11), 2415-2430 (2015).

37. Zgrabik, C. M., & Hu, E. L. Optimization of sputtered titanium nitride as a tunable metal for plasmonic applications. *Optical Materials Express*, **5**(12), 2786-2797 (2015).

38. Palik E. Handbook of Optical Constants of Solids, Vol. 2, Ed., Academic Press, New York (1991).

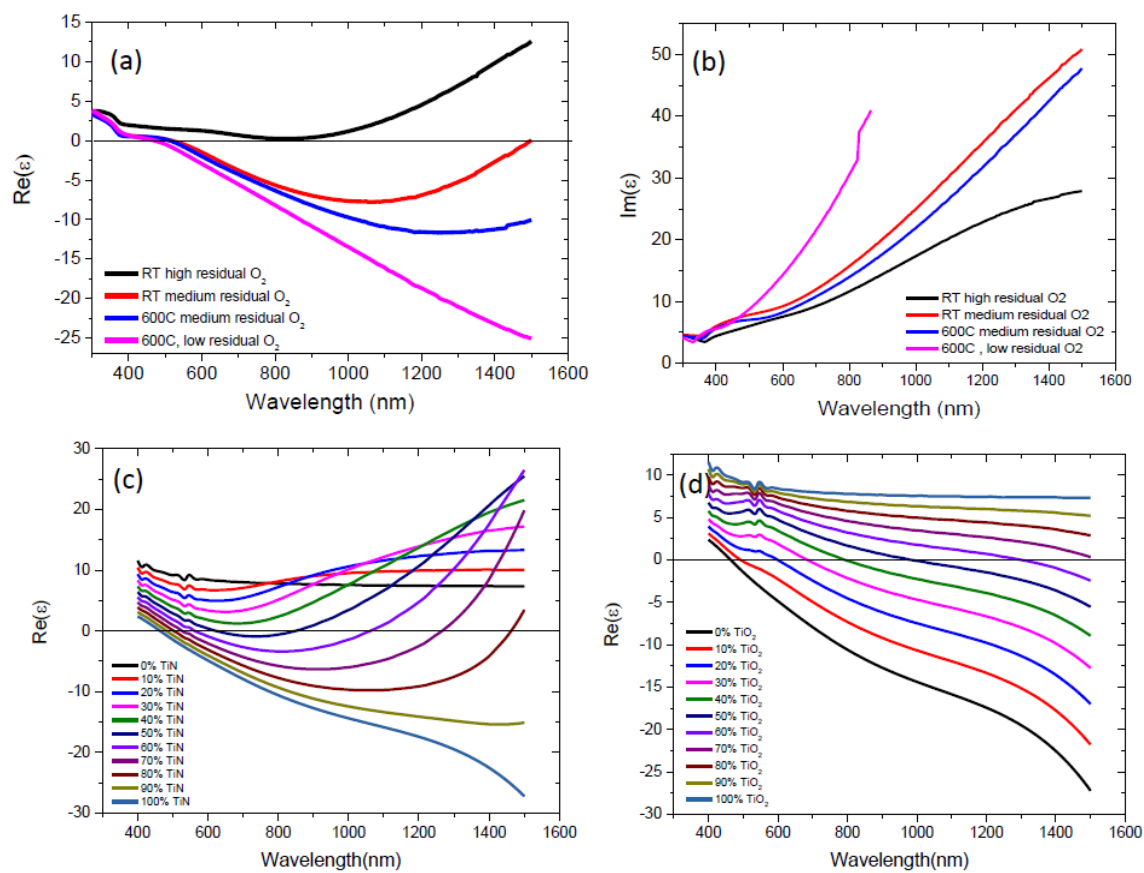

**Figure 1**. (a) Real and (b) imaginary dielectric constants of the films deposited at the different partial oxygen content and temperatures as indicated in the panels. (c) and (d) Dielectric constants simulated using a Maxwell-Garnett model considering a composite formed by TiN inclusions in TiO$_2$ (c) and TiO$_2$ inclusions in TiN (d).

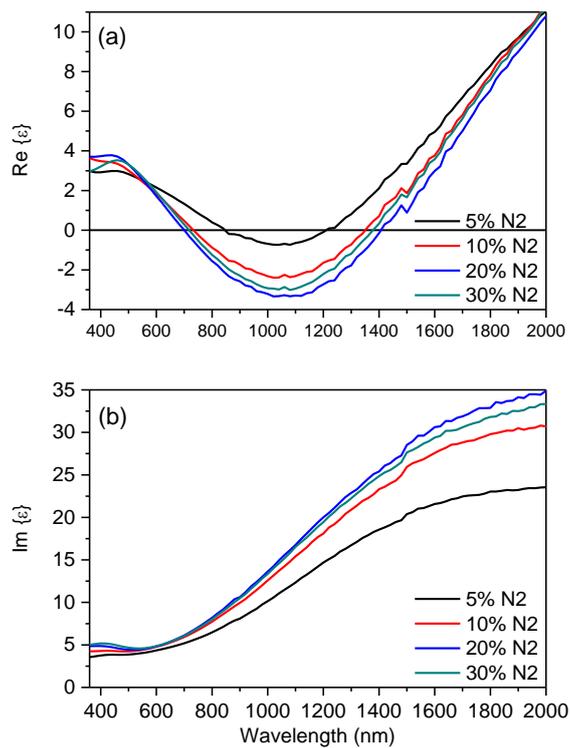

**Figure 2.** The real (a) and imaginary (b) dielectric constants of the RT fabricated films under different N$_2$ partial pressures.

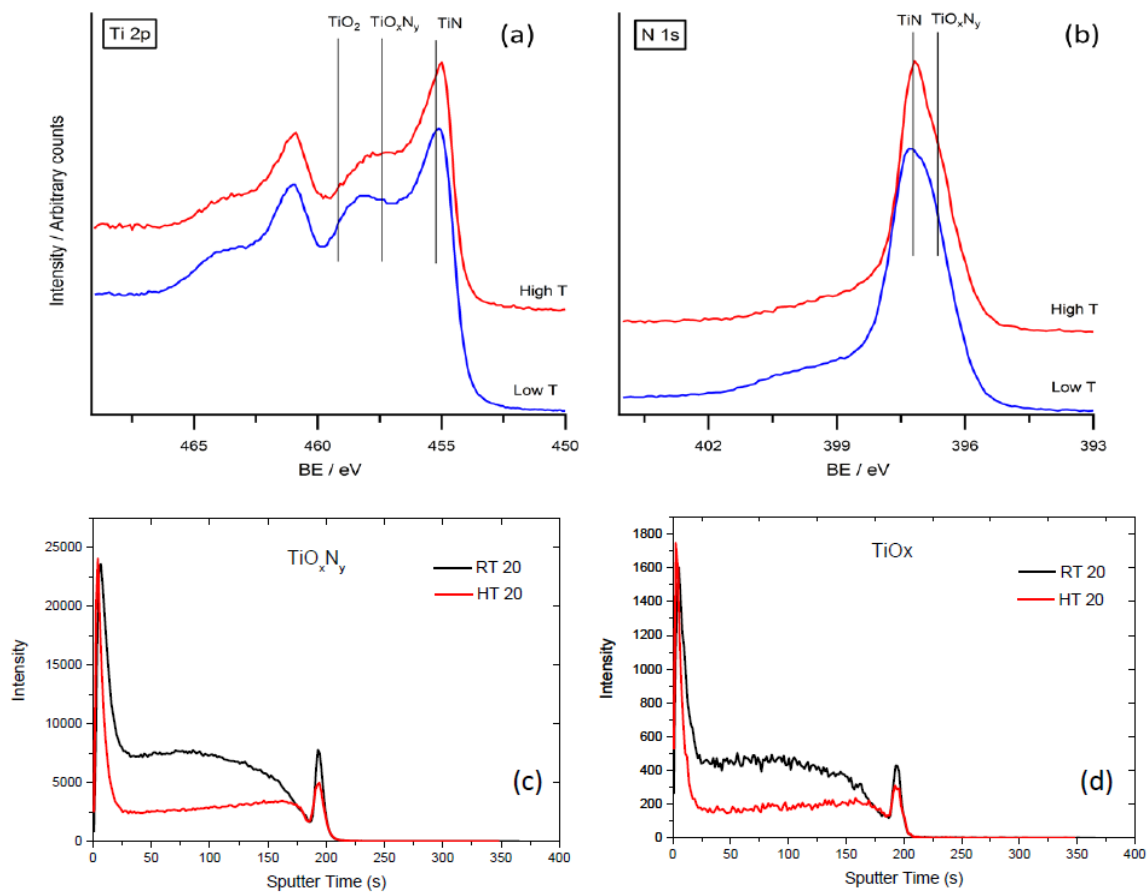

**Figure 3.** (a) Ti 2p XP spectra of samples deposited at 20% N$_2$ partial pressure for 600C and room temperature after a few consecutive sputtering cycles. (b) N 1s spectra for the same conditions. The lines mark characteristic binding energies of different Ti chemical states. (c) and (d) SIMS profiles of 20% Nitrogen for TiOxNy (c) and TiOx (d) phases.

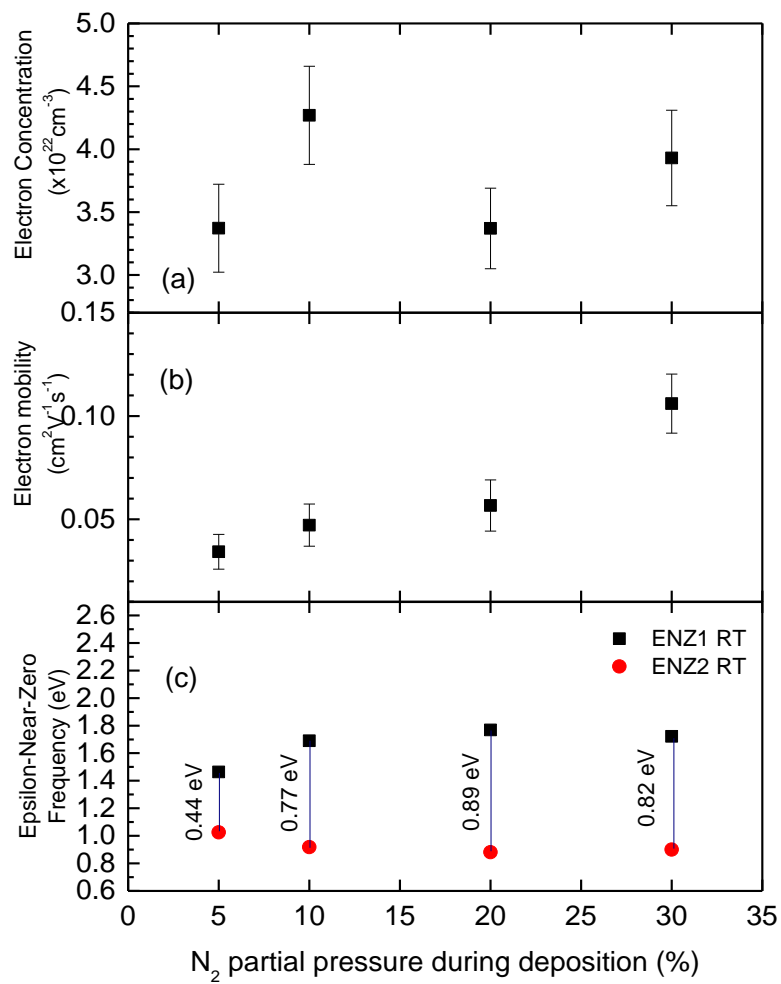

**Figure 4.** a) Carrier concentration, b) carrier mobility and c) ENZ band tuning versus Nitrogen partial pressure for medium oxygen content films.

**Table 1.** Resistance, resistivity and partial pressure before sputtering for the three chamber preparation scenarios studied

| Deposition Temperature | Pre-deposition conditioning | Partial oxygen pressure (Torr) | Resistance ($\Omega$) | Resistivity ($\Omega$m $\times 10^{-6}$) |
|---|---|---|---|---|
| Room Temperature | None | 2.80E-08 | 47 | 9.2 |
| Room Temperature | Preheating at 250 | 1.80E-08 | 14 | 2.5 |
| 600C | Heated at 600 and Ti pre-sputter | 7.00E-09 | 10 | 2.0 |

**Table 2:** Chemical binding of Ti in the samples fabricated at 20% $N_2$ partial pressure.

| Sample deposition temperature | TiN (%) | TiO$_x$N$_y$ (%) | TiO$_2$ (%) |
|---|---|---|---|
| 600 °C | 87 | 9.6 | 3.5 |
| RT | 71.3 | 18.6 | 10.1 |